%% file: supercurrent_fi.tex
\documentclass[11pt]{article}

\usepackage{slashed,amssymb}

\include{sugracom}

\newcommand{\bphi}{{\bar\phi}}
\newcommand{\bPhi}{{\bar\Phi}}

\newcommand{\bpsi}{{\bar\psi}}

\newcommand{\bj}{{\bar j}}

\newcommand{\eol}{\notag \\}

\usepackage{amsmath}
\usepackage{amsfonts}
\usepackage{wasysym}
\usepackage{graphicx}

\usepackage{cite}

\usepackage{hyperref}

\numberwithin{equation}{section}

\begin{document}
\thispagestyle{empty}


\hfill UCB-PTH-10/05

\hfill arXiv:1003.0249 [hep-th]

\hfill March 2, 2010

\addvspace{45pt}

\begin{center}

\Large{\textbf{Conserved supercurrents and Fayet-Iliopoulos terms in supergravity}}
\\[35pt]
\large{Daniel Butter}
\\[10pt]
\textit{Department of Physics, University of California, Berkeley}
\\ \textit{and}
\\ \textit{Theoretical Physics Group, Lawrence Berkeley National Laboratory}
\\ \textit{Berkeley, CA 94720, USA}
\\[10pt] 
dbutter@berkeley.edu
\end{center}

\addvspace{35pt}

\begin{abstract}
\noindent
Recently there has appeared in the literature a sequence of papers questioning
the consistency of supergravity coupled to Fayet-Iliopoulos
terms. A key feature of these arguments is a demonstration that the
conventional superspace stress tensor fails to be gauge invariant.
We briefly show here how this can be understood as defining the stress
tensor in a non-covariant Brans-Dicke frame in an underlying superconformal
theory. When converted to the Einstein frame, the inconsistency vanishes,
which is consistent with the emergence of a global symmetry discussed in these
papers.
\end{abstract}

\setcounter{tocdepth}{2}
\newpage
\setcounter{page}{1}

\section{Introduction}
Globally supersymmetric theories admit a peculiar type of term, known as a
Fayet-Iliopoulos D-term, which in superspace language may be written
\begin{align}
S_{FI} = 2 \xi \int d^8z \, V = -\xi \int d^4x\, D
\end{align}
up to a total derivative (which we shall always discard).
Here $V$ is the gauge prepotential of some $U(1)$ (denote it $U(1)_\xi$),
$D = -D^\alpha \bar D^2 D_\alpha V / 8$ is its highest component,
and $\xi$ is a parameter of dimension two. The gauge
invariance of this expression follows since under a gauge transformation
\[
V \rightarrow V + \Lambda + \bar\Lambda
\]
for chiral $\Lambda$,
\begin{align}\label{eq_temp1}
\int d^8z \, \Lambda = 0.
\end{align}
The coupling of such terms to conventional (old minimal) supergravity is not
entirely straightforward since the direct analogue of \eqref{eq_temp1} fails to be true
there, but a construction exists which we will
review shortly.\footnote{See the extensive citations in \cite{Komargodski:2009pc} 
and \cite{Dienes:2009td} on the subject.}

Recently it has been argued by Komargodski and Seiberg \cite{Komargodski:2009pc}
that the inclusion of such a term in supergravity is problematic: specifically,
its contribution to the superspace stress tensor of supergravity fails to be gauge
invariant at the classical level without requiring a global
symmetry.\footnote{Within the last two weeks, Komargodski and
Seiberg have released another paper \cite{Komargodski:2010rb} constructing
an alternative supercurrent which appears to correspond to a non-minimal
16+16 supergravity in which the FI term is more easily accounted for. This seems
to correspond to adding both a chiral and linear superfield to conformal supergravity
to yield the non-minimal Poincar\'e sector.}  This has been expounded upon in detail by
Dienes and Thomas \cite{Dienes:2009td}, who examined the situation
in further detail in both old and new minimal supergravity, and by
Kuzenko \cite{Kuzenko:2009ym}, who clarified several issues relating to
the existence of the supercurrents.

We would like to comment on some of the issues involved in this argument
by considering the full coupling of the theory to supergravity in superspace,
specifically, the super-Weyl redefinitions which must accompany any coupling of a D-term in conventional
(old minimal) supergravity. The easiest way to see the issue is to
consider not an actual Fayet-Iliopoulos term, but another term rather closely
related and even more mundane: the K\"ahler potential. Recall that a
supersymmetric non-linear sigma model can be constructed from the D-term
\begin{align}
\int d^8z\, K = \int d^4x \left(-K_{i\bj} \partial^m \phi^i \partial_m \bphi^\bj + \ldots \right)
\end{align}
On the right hand side of this formula only the K\"ahler metric $K_{i\bj}$ (and
its derivatives) appears; the reason is that the left hand side is invariant
under $K \rightarrow K + F + \bar F$ for chiral $F$ and so must be the right side.
(In fact, the K\"ahler potential term may be understood as a composite
Fayet-Iliopoulos term for some $U(1)_K$.)

Naively coupling such a term to supergravity is a bit problematic.
Students of conventional (old minimal) supergravity are well-aware of a certain
curious feature: the integral of the supervolume is proportional to the supersymmetric
Einstein-Hilbert term. Specifically
\begin{align} \label{eq_EH}
-\frac{3}{\kappa^2} \int d^8z\,E = S_{EH} = -\frac{1}{2 \kappa^2} \int d^4x\sqrt{g}\, \mathcal R + \ldots
\end{align}
where $\kappa^2$ is the reduced Planck length.
This particularly causes problems when coupling a K\"ahler potential to
supergravity. The coupling one would naively propose is $\int d^4\theta\, E\, K$;
however, it is clear that such a term must yield a non-canonical Einstein-Hilbert term
\begin{align} \label{eq_Kahler}
\int d^8z \,E\, K = \int d^4x\sqrt{g}\,
     \left(\frac{1}{6} K \mathcal R - K_{i \bj} \nabla^m \phi^i \nabla_m \bphi^\bj + \ldots \right)
\end{align}
as the previous formula is simply a special case of constant $K$. This Brans-Dicke
like interaction can be cured by performing a super-Weyl transformation on
the supergravity sector, and one finds that the proper way to couple
supergravity to chiral matter is via an exponential and a super-Weyl
redefinition \cite{wb}
\begin{align}\label{eq_EHKold}
-\frac{3}{\kappa^2} \int d^8z\, E\, e^{-\kappa^2 K / 3}
     &= \int d^4x \sqrt{g} \, e^{-\kappa^2 K/3} \left(-\frac{1}{2 \kappa^2} \mathcal R - K_{i \bj} \nabla^m \phi^i \nabla_m \bphi^\bj + \ldots \right) \\ \label{eq_EHK}
     &= \int d^4x \sqrt{g'} \left(-\frac{1}{2 \kappa^2} \mathcal R' - K_{i \bj} \nabla^m \phi^i \nabla_m \bphi^\bj + \ldots \right)
\end{align}
The first equality, \eqref{eq_EHKold}, is supergravity coupled to chiral matter
in what we shall call the Brans-Dicke frame; the second equality, \eqref{eq_EHK},
is in the Einstein frame. An important feature of the right-hand side of either
of these formulae is that there is no clear
distinction between a pure supergravity term and a chiral matter term; rather, they
are blended together. This is obvious for \eqref{eq_EHKold}
but it is also true for \eqref{eq_EHK} since after the super-Weyl transformations,
the supersymmetry transformation rule for the gravitino includes details of the matter
sector (albeit suppressed by factors of $\kappa^2$). In fact, the entire artifice of
K\"ahler superspace has been worked out to
explain the details of this intertwining \cite{bgg}. One important detail
that we should keep in mind is that \eqref{eq_EHKold} is not K\"ahler invariant
without performing additional super-Weyl transformations whereas \eqref{eq_EHK}
is. Classically then we should regard \eqref{eq_EHK} as the proper frame in
which to perform our calculations if we would like K\"ahler invariance to be
preserved. Indeed, this is what one normally does \cite{wb}.

Moreover, and this is the critical part, the two equations \eqref{eq_EHKold} and
\eqref{eq_EHK} \emph{differ in their form even in the small $\kappa^2$ limit}. Even though
we have suppressed all but two terms, these alone clearly differ
in their form by $K \mathcal R /6$, which survives in the small $\kappa^2$ limit.
Although this term certainly vanishes when one turns off supergravity, we
find the stress tensor by varying the metric and \emph{then}
setting it to zero; thus terms linear in the curvature can indeed alter
the stress tensor. This one, for example, contributes to the stress tensor
a term $\partial_m \partial_n K$ which is clearly not K\"ahler
invariant. Thus, the canonical definition of the stress tensor can (and does!) differ between
the Brans-Dicke and the Einstein frames.

The coupling of an FI term to supergravity is equivalent to that of $K$.
One can simply make the replacement of $K \rightarrow 2\xi V$ in the above argument,
and quite analogously we expect that the definition of the stress tensor should differ between
the two frames. Our contention is that the specific calculation recounted 
in \cite{Komargodski:2009pc, Dienes:2009td} while performed in the rigid limit
is ultimately equivalent to a calculation performed in the superspace analogue of the
Brans-Dicke frame; since that frame
fails to be gauge invariant it is unsurprising that its supercurrent should have the
same problem. We will further show that the superspace Einstein frame possesses
a gauge invariant stress tensor, but implies the additional symmetry
these authors discussed, reaffirming their main point.

Our paper is structured as follows. In section 2, we review the calculation
of the gauge supercurrent in a globally supersymmetric theory both as a 
warmup and to set our conventions. In section 3, we consider the supergravity
supercurrent, first in the superspace context where we discuss the difference
between the Brans-Dicke and Einstein frames, and then in the component
context where we demonstrate that the Einstein frame currents do give the
standard globally supersymmetric currents when supergravity is decoupled.
An appendix is attached which briefly summarizes the calculation of the
Noether currents.

\section{Review: The gauge supercurrent and some conventions}\label{review_gauge}
Before diving into the details of supergravity, we will briefly discuss gauge
supercurrents both to standardize our gauge superfield notation and to review how
superspace currents yield the more familiar component currents.

The action of a $U(1)$ gauge sector with a Fayet-Iliopoulos term
coupled to a single chiral superfield  of charge $g$ in global supersymmetry
may be written in superspace as
\begin{align}\label{eq_flatSYM}
S = \int d^8z\, \left(\bar\phi \, e^{2g V} \phi + 2\xi V\right)
     + \frac{1}{4} \int d^6z\, W^\alpha W_\alpha
     + \frac{1}{4} \int d^6\bar z\, \bar W_\dalpha \bar W^\dalpha
\end{align}
The chiral superfield $\phi$ is (conventionally) chiral, obeying $D^\dalpha \phi = 0$,
as is the $U(1)$ gaugino field strength. The gauge invariance of the action follows
from the transformations
\begin{gather}\label{eq_gaugetfs}
\phi \rightarrow e^{-2g\Lambda} \phi, \;\;\;
\bar\phi \rightarrow \bar\phi \, e^{-2g\bar\Lambda}, \;\;\;
V \rightarrow V + \Lambda + \bar\Lambda
\end{gather}
where $\Lambda$ is also chiral. (Normal gauge transformations correspond
to $\lambda = i (\Lambda - \bar\Lambda)$.) In the conventions we will use here,
the gaugino field strength is defined in terms of the $U(1)$ prepotential $V$ as
\begin{align}
W_\alpha = \frac{1}{4} \bar D^2 D_\alpha V, \;\;\;
\bar W^\dalpha = \frac{1}{4} D^2 \bar D^\dalpha V
\end{align}

The fundamental dynamical variables of this theory are $\phi$, $\bar\phi$, and $V$.
If we vary the action under small deformations of each of these parameters,
\begin{gather}
\delta V \equiv \Sigma, \;\;\;
\delta \phi \equiv \eta, \;\;\;
\delta \bphi \equiv \bar\eta
\end{gather}
we find a general first order structure
\begin{align}\label{eq_g1super}
S^{(1)} = \int d^8z \, \Sigma J_V
     + \int d^6z\, \eta\, J_\phi
     + \int d^6\bar z\, \bar \eta\, \bar J_\phi
\end{align}
where for our simple model
\begin{gather}
J_V = 2\xi + D^\alpha W_\alpha + 2g \bar\phi e^{2gV} \phi \\
J_\phi = -\frac{1}{4} \bar D^2 (\bar\phi e^{2gV}) \\
\bar J_\phi = -\frac{1}{4} D^2 (e^{2gV} \phi)
\end{gather}
Note that the currents have properties similar to the first order
variations to which they couple: $J_V$ is real, $J_\phi$ is chiral, and
$\bar J_\phi$ is antichiral.

If the original action is gauge invariant, then $S^{(1)}$ should
vanish when we choose the deformations of the fields to be
gauge transformations. Thus for infinitestimal $\Lambda$ we
find the general structure
\begin{align}
0 = \delta_g S  = \int d^6z \, \Lambda \left(-\frac{1}{4} \bar D^2 J_V - 2g \phi J_\phi\right)
     + \hc
\end{align}
implying the classical conservation equations
\begin{align}\label{eq_gsuperflat}
-\frac{1}{4} \bar D^2 J_V = 2g \phi J_\phi, \;\;\;
-\frac{1}{4} D^2 J_V = 2g \bphi \bar J_\bphi
\end{align}
When the matter fields are placed on shell, $J_\phi$ and $\bar J_\phi$ vanish
and this becomes the usual superfield version of current conservation.

To better see this, we can turn to a component formulation of the same theory.
The original way of doing this is to go to Wess-Zumino gauge for the prepotential
$V$, but there is a more elegant geometric approach. One promotes
the conventionally chiral superfield $\phi$ to a covariantly
chiral $\Phi$ by endowing the superspace derivatives with a gauge connection.
In this language
\[
\bphi e^{2gV} \phi \rightarrow \bPhi \Phi, \;\;\;\;
D^2 (e^{2gV} \phi) \rightarrow \CD^2 \Phi, \;\;\;\;
\bar D^2 (\bar\phi e^{2gV}) \rightarrow \BCD^2 \bar\Phi
\]
and except for the explicit FI term, the prepotential $V$ need not be mentioned
explicitly.\footnote{See the classic textbooks \cite{superspace, Buchbinder:1998qv}
for a discussion of this.} We then define the covariant combinations
\[
\chi_\alpha = \frac{1}{\sqrt 2} \CD_\alpha \Phi, \;\;\;
F = -\frac{1}{4} \CD^2 \Phi
\]
for the matter sector. For the gauge sector, we have
\begin{gather}
A_m = -\frac{1}{4} \bsigma_m^{\dalpha \alpha} [\CD_\alpha, \BCD_\dalpha] V \equiv -\Delta_m V \\
\lambda_\alpha = W_\alpha, \;\;\;
D = -\frac{1}{2} \CD^\alpha W_\alpha
\end{gather}
We have introduced the notation $\Delta_{\alpha \dalpha} \equiv -[\CD_\alpha, \BCD_\dalpha]/2$
for selecting out the vector component (i.e. the component of $\theta \sigma^m \btheta$
in the superfield expansion). Writing \eqref{eq_flatSYM} in component notation gives
\begin{align}
\Lag =& -\CD^m \bphi \CD_m \phi - i \bar\chi \bsigma^m \CD_m \chi + \bar F F
     + \sqrt{2} g (\lambda \chi) \bphi + \sqrt{2} g (\bar\lambda \bar\chi) \phi
     - g D \phi \phi\eol
     & - \frac{1}{4} F^{mn} F_{mn} - i \bar \lambda \bsigma^m \CD_m \lambda + \frac{1}{2} D^2 - \xi D
\end{align}
Its first order variation has the general form
\begin{align}
S^{(1)} = \int d^4x\, \biggl(
     & \delta \phi J_{\phi}^{(\phi)}
     + \delta \chi^{\alpha} J_{\phi \alpha}
     + \delta F J_\phi + \hc \eol
     & + \delta A_m{} J_V^m
     + \delta \lambda^{\alpha} J_{V \alpha}
     + \delta \bar\lambda_\dalpha J^\dalpha_{V}
     + \delta D J_V^{(D)} \biggr)
\end{align}
For our specific case, it is easy to work out the various currents $J$ in the above
expression, but a more profitable approach is to identify them from the component
version of \eqref{eq_g1super}. Doing so gives
\begin{gather}
J_\phi^{(\phi)} = - \frac{1}{4} \CD^2 J_\phi \\
J_{\phi \alpha} = -\frac{1}{\sqrt 2} \CD_\alpha J_\phi
\end{gather}
for the matter supermultiplet currents, and
\begin{gather}
J_{V\alpha} = \frac{1}{2} \CD_\alpha J_V,  \;\;\;
J_V^\dalpha = \frac{1}{2} \BCD^\dalpha J_V \\
J_V^m = \frac{1}{2} \Delta^m J_V \\
J_V^{(D)} = -\frac{1}{2} J_V
\end{gather}
for the gauge supermultiplet currents.

Invariance of the component action under gauge transformations implies
\begin{align}\label{eq_gcompflat}
\partial_m J_V{}^m =\,
     & i g\phi J^{(\phi)}_\phi
     + ig \chi^{\alpha} J_{\phi\alpha}
     + ig F J_\phi + \hc
\end{align}
It is easy to see that \eqref{eq_gcompflat} is a consequence of
\eqref{eq_gsuperflat}: the latter implies
\begin{align}
\frac{i}{32} [\CD^2, \BCD^2] J_V = -\frac{i}{8} \CD^2 (2g \phi J_\phi) + \hc
\end{align}
which reduces to the former after some algebra.

Turning off the gauge sector amounts to setting $V=0$, which yields in our simple case the
current superfield
\[
J_V = 2\xi + 2g \bar\phi \phi
\]
whose vector component
\[
J_V{}^m = i g \partial^m \bphi \phi - i g \bphi\partial^m \phi - g (\bar\chi \sigma^m \chi)
\]
is precisely what would have been constructed by the Noether procedure.
This is an elegant (if historically backward) approach to constructing Noether currents
associated with gauge fields: vary the action with respect to the gauge
field and then set it to zero. It is this approach which is most easily
replicated in superspace to construct the supergravity supercurrents.

\section{The supergravity supercurrent}
We turn now to our actual object of interest, the supergravity supercurrent.
In order to understand the superfield form of the current equations we will
derive, it helps to recast the original theory in a superconformal
form.\footnote{The importance of superconformal methods in understanding
the occasionally bizarre structure of conventional supergravity cannot be overstated.
For the construction of superspace with the full superconformal algebra as the
structure group, see the discussion in \cite{butter1}.} We will briefly review why.

\subsection{The relevance of superconformal concerns}
Recall that when the bosonic conformal algebra is placed in the structure group
of a regular Poincar\'e theory, the covariant d'Alembertian receives a correction proportional
to the Ricci scalar when acting on a scalar fields $\phi$ of conformal dimension one:
\[
\nabla^a \nabla_a \phi = \CD^a \CD_a \phi + \frac{1}{6} \mathcal R\, \phi
\]
In the above, $\nabla$ denotes the covariant derivative with the conformal algebra
in the structure group while $\CD$ denotes the normal Poincar\'e derivative.
The conformally invariant kinetic term for a scalar, strangely normalized with
a factor of $-3$ contains a Brans-Dicke like term:
\[
-3 \int d^4x \sqrt{g}\, \phi \nabla^a \nabla_a \phi = \int d^4x \sqrt{g}\, \left(-3\phi \CD^a \CD_a \phi
     - \frac{1}{2} \mathcal R \phi^2 \right)
\]
The field $\phi$ is known as the ``conformal compensator,''
and taking the conformal gauge $\phi=1$ then yields the conventional Einstein-Hilbert
Lagrangian.

A similar approach ``explains'' the bizarre features of conventional supergravity.
Introducing a chiral superfield $\phi_0$ with conformal dimension one, the kinetic
term (again with an additional factor of $-3$) is
\begin{align}
-\frac{3}{\kappa^2} \int d^4\theta\, E\, \bar\phi_0 \phi_0 = \int d^4x \sqrt{g}\,
     \left(-\frac{3}{\kappa^2} \bar\phi_0 \CD^a \CD_a \phi_0
          - \frac{1}{2\kappa^2} \mathcal R \abs{\phi_0}^2 + \ldots \right)
\end{align}
The coupling of a K\"ahler potential to this theory is that of a composite $U(1)$ factor:
\begin{align}
-\frac{3}{\kappa^2} \int d^4\theta\, E\, \bar\phi_0 \phi_0\, e^{-\kappa^2 K/3}
     = \int d^4x \sqrt{g}\, \biggl(&
     -\frac{3}{\kappa^2} e^{-\kappa^2 K/3} \bar\phi_0 \nabla^a \nabla_a \phi_0
     + \frac{1}{16}\abs{\phi_0^2} \nabla^\alpha \bar\nabla^2 \nabla_\alpha K + \ldots \biggr) \eol
     = \int d^4x \sqrt{g}\, \biggl(&
     -\frac{3}{\kappa^2} e^{-\kappa^2 K/3} \bar\phi_0 \CD^a \CD_a \phi_0
     - \frac{1}{2} e^{-\kappa^2 K/3} \mathcal R \abs{\phi_0}^2\eol
     & + \abs{\phi_0}^2 K_{i \bj} \nabla^m \phi^I \nabla_m \bphi^\bj + \ldots \biggr)
\end{align}
Indeed, the part of the action corresponding to the conventional K\"ahler
action is simply the $D$-term of the composite $U(1)_K$. 

It is clear that the a certain superconformal gauge choice ($\phi=1$) must correspond
to the Brans-Dicke frame \eqref{eq_EHKold} while another ($\phi = e^{\kappa^2 K/6}$)
must correspond to the Einstein frame \eqref{eq_EHK}. The advantage of working
in a superconformal framework is that we can impose these gauge choices at the
superfield level \emph{without} first going to components.
This will be critical in finding the Einstein frame superspace supercurrents.

It was shown in \cite{butter1} that one can reduce conformal superspace
to Poincar\'e superspace with a residual $U(1)$ structure on which the conformal
transformations are realized non-linearly. The resulting Poincar\'e $U(1)$ superspace
(which we will refer to simply as ``Poincar\'e'' from now on) is that of
\cite{Muller:1985vg}. It is not necessary for the $U(1)$ to actually contain
degrees of freedom; one can, for example, rewrite the original
supergravity of \cite{wb} in this structure, so it is quite
generic. In \cite{bgg}, it was shown how to use the $U(1)$ structure to
encode K\"ahler transformations, but the discussion there can easily
be generalized to include a $U(1)_\xi$ rather than a K\"ahler potential.
It is this structure we eventually expect to recover in the superspace
currents.

\subsection{The superspace conformal and Poincar\'e stress tensors}
We will use the convenient shorthands
\begin{gather}
\left[Z \right]_D \equiv \int d^4x \, d^4\theta\, E\, Z, \;\;\;
\left[W\right]_F \equiv \int d^4x \, d^2\theta\, \chE\, W
\end{gather}
to denote integrations over the full superspace with integrand $Z$
and over the chiral superspace with integrand $W$, respectively.
For now we will maintain manifest superconformal
invariance at all times; thus the above terms are invariant only if
$Z$ is conformally primary with conformal dimension two and vanishing
$U(1)_R$ weight and if $W$ is conformally primary and chiral with conformal
dimension three and $U(1)_R$ weight two. (If a superfield has
conformal dimension $\Delta$ and $U(1)_R$ weight $w$, we will
refer to its weight as $(\Delta, w)$; thus $Z$ is weight $(2,0)$
and $W$ is weight $(3,2)$.) $E$ denotes the superdeterminant
of the supervierbein, a suitable volume measure for the full superspace,
while $\chE$ is the chiral volume measure \cite{butter1}.

We are interested ultimately in examining the classical supercurrent for the
D-term action
\begin{align}
S = -\frac{3}{\kappa^2} \biggl[\phi_0 \bar\phi_0 \, e^{-2\kappa^2 \xi V/3} \biggr]_D
     + \biggl[\frac{1}{4} W^\alpha W_\alpha\biggr]_F
     + \biggl[\frac{1}{4} W_\dalpha W^\dalpha \biggr]_F
\end{align}
where $W_\alpha$ is the gaugino superfield and $V$ the prepotential
associated with the $U(1)_\xi$.
It is convenient to switch from conventional chirality to covariant
chirality for the purposes of making contact with the notation used in
\cite{butter2}:
\begin{align}
S = -\frac{3}{\kappa^2} \biggl[\Phi_0 \bar\Phi_0\biggr]_D
     + \biggl[\frac{1}{4} W^\alpha W_\alpha \biggr]_F
     + \biggl[\frac{1}{4} W_\dalpha W^\dalpha \biggr]_{\bar F}
\end{align}
where $\Phi_0$ is covariantly chiral with a $U(1)_\xi$ charge
of $-\kappa^2 \xi / 3$. The first order variation of this model has the
form\footnote{In \cite{butter2}, we used $V_a$ to denote $H_a$. Here we
use the latter notation to agree with the general convention.}
\begin{align}\label{eq_super1var}
S^{(1)} = \biggl[H^{\dalpha \alpha} J_{\alpha \dalpha} + \Sigma J_\xi \biggr]_D
     + \biggl[\eta_0 J_0\biggr]_F
     + \biggl[\bar\eta_0 \bar J_0\biggr]_{\bar F}
\end{align}
where $H_a$ is an Hermitian superfield whose components encode
the variation of the conformal supergravity multiplet, $\Sigma$ is
a Hermitian superfield encoding the variation of the $U(1)_\xi$ gauge
multiplet, and $\eta_0$ is a chiral superfield encoding the variation of the
chiral compensator. More precise definitions of the
variational superfields are given in \cite{butter2}.\footnote{In particular, the definition
of $\Sigma$ differs from the conventional definition by terms involving $H_a$
which render its transformation rule covariant. The question of which (if either)
definition is more correct is moot since we will soon place the gauge superfield on
shell.}

Of immediate concern are the superconformal properties of the various
objects. $H_a$ is weight $(-1,0)$ and so $J_a$ must be weight $(3,0)$.
$\Sigma$ is weight $(0,0)$ and thus $J_\xi$ is weight $(2,0)$.
Note that $\eta_0$ is necessarily of weight $(1,2/3)$ and so $J_0$ has
weight $(2, 4/3)$. Also of importance is that $J_0$ has $U(1)_\xi$ charge of
$+\kappa^2 \xi / 3$, opposite that of $\Phi_0$.

Under a (quantum) gauge transformation\footnote{Recall that when a theory with a gauge
invariance is expanded in terms of first order quantum variations about a background,
there exist two different notions of gauge transformation: background
transformations under which the quantum variations transform homogeously
and quantum transformations under which the background is invariant. The
latter are important for figuring out the currents.}
\begin{gather*}
\delta \Sigma = \Lambda + \bar\Lambda, \;\;\;
\delta \eta_0 = \frac{2}{3} \kappa^2 \xi \Lambda \Phi_0, \;\;\;
\delta \bar\eta_0 = \frac{2}{3} \kappa^2 \xi \bar \Lambda \bPhi_0
\end{gather*}
Gauge invariance of the first-order action implies that\footnotemark
\begin{align}\label{eq_gsuper}
\CP J_\xi = -\frac{2}{3} \kappa^2 \xi \Phi_0 J_0, \;\;\;
\ACP J_\xi = -\frac{2}{3} \kappa^2 \xi \bPhi_0 J_0
\end{align}
where $\CP \equiv -\bar\nabla^2 / 4$ and $\ACP \equiv -\nabla^2/4$.
\footnotetext{In conformal superspace, these are chiral projection operators, just as they
are in global supersymmetry, and so $\CP J_\xi$ is a chiral superfield.}
This is the superfield version of gauge current conservation which we
have discussed previously in the globally supersymmetric case.

A similar structure exists for conformal supergravity transformations.
Under these,
\begin{gather}
\delta H_{\alpha \dalpha} = \nabla_\alpha L_\dalpha - \nabla_\dalpha L_\alpha, \;\;\;
\delta \Sigma = L^\alpha W_\alpha + L_\dalpha W^\dalpha \eol
\delta \eta_0 = \CP(L^\alpha \nabla_\alpha \Phi_0)
     + \frac{1}{3} \Phi_0 \,\CP \nabla^\alpha L_\alpha, \;\;\;
\delta \bar\eta_0 = \ACP(L_\dalpha \nabla^\dalpha \bPhi_0)
     + \frac{1}{3} \bPhi_0\,\ACP \nabla_\dalpha L^\dalpha
\end{gather}
which imply the superfield version of energy-momentum conservation
\begin{gather}
\nabla^\dalpha J_{\alpha \dalpha} = + W_\alpha{} J_\xi
     + \nabla_\alpha \Phi_0 J_0
     - \frac{1}{3} \nabla_\alpha (\Phi_0 J_0) \eol
\nabla^\alpha J_{\alpha \dalpha} = + W_\dalpha{} J_\xi
     + \nabla_\dalpha \bPhi_0 \bar J_0
     - \frac{1}{3} \nabla_\dalpha (\bPhi_0 \bar J_0)
\end{gather}

Using the techniques developed in \cite{butter2}, one can show
that for the model under consideration here,\footnote{$W_\alpha$ here differs by a factor of $i$ from that
defined in \cite{butter2}.}
\begin{gather}\label{eq_cstress}
J_{\alpha \dalpha} = \frac{2}{\kappa^2} X \hat G_{\alpha \dalpha} + W_\alpha W_\dalpha \\
J_0 = -\frac{6}{\kappa^2 \Phi_0} X \hat R \\
J_\xi = 2 \xi X + \nabla^\alpha W_\alpha
\end{gather}
where we have defined
\begin{gather}
X \equiv \Phi_0 \bar\Phi_0 \\
\hat G_{\alpha \dalpha} \equiv - X^{1/2} \Delta_{\alpha \dalpha} X^{-1/2} \\
\hat R \equiv -\frac{1}{8} X^{-1} \bar\nabla^2 X
\end{gather}
The superfields $\hat G_{\alpha \dalpha}$ and $\hat R$ are superconformally
primary when $X$ is of dimension two and reduce to the Poincar\'e superfields
of the same name when the gauge choice $X=1$ is made.

We now have the formulae governing the conservation of energy and
momentum in superspace. In order to find the Poincar\'e stress energy relation,
we first put the gauge sector on shell (i.e. we set $J_\xi$ to zero) and then
choose the conformal gauge. There are essentially two options available to us.

\subsubsection{The Brans-Dicke frame}
The Brans-Dicke frame corresponds to choosing the \emph{conventionally chiral}
superfield $\phi_0$ to be unity. This is equivalent to choosing
\[
X = \bphi_0 \phi_0 e^{-2 \xi \kappa^2 V / 3} \rightarrow e^{-2 \xi \kappa^2 V / 3}
\]
in the original action and thus corresponds (in the small $\kappa^2$ limit)
to the choice made in \cite{Komargodski:2009pc, Dienes:2009td}.
The easiest way to see this is to rewrite the superfields in conventionally chiral
notation (that is, remove $U(1)_\xi$ from the structure group)
and then to set $\phi_0=1$, fixing the $U(1)_R$ gauge along with the conformal
symmetry. The superfields $\hat G_{\alpha \dalpha}$ and $\hat R$ that we have previously
defined become
\begin{gather}
\hat G_{\alpha \dalpha} = G - \frac{1}{3} \xi \kappa^2 \Delta_{\alpha \dalpha} V
     + \mathcal O(\kappa^4) \\
\hat R = R + \frac{1}{12} \xi \kappa^2 \BCD^2 V + \mathcal O(\kappa^4)
\end{gather}
where the Brans-Dicke frame superfields $G$ and $R$ obey the constraint
\begin{align}
\CD^\dalpha G_{\alpha \dalpha} = \CD_\alpha R
\end{align}
We can rewrite the conservation equation as
\begin{align}\label{eq_temp4}
\CD^\dalpha J_{\alpha \dalpha} = -\frac{2}{3} \kappa^2 \xi \CD_\alpha V (\Phi_0 J_0)
     - \frac{1}{3} \CD_\alpha (\Phi_0 J_0)
\end{align}
where we have left the gauge invariant combination $\Phi_0 J_0$ in covariant form.
(Note that since $\Phi_0 J_0$ is gauge invariant, it is both covariantly \emph{and}
conventionally chiral.) It expands out as
\begin{align}\label{eq_temp3}
\Phi_0 J_0 = -\frac{6}{\kappa^2} e^{-2\kappa^2 \xi V/3} \hat R
     = -\frac{6}{\kappa^2} R + 4 \xi V R - \frac{1}{2} \xi \BCD^2 V + \mathcal O(\kappa^2)
\end{align}
The supercurrent is 
\begin{gather}\label{eq_temp2}
J_{\alpha \dalpha} = \frac{2}{\kappa^2} G_{\alpha \dalpha}
     - \frac{4}{3} \xi V G_{\alpha \dalpha}
     - \frac{2}{3} \xi \Delta_{\alpha \dalpha} V + W_\alpha W_\dalpha
     + \mathcal O(\kappa^2)
\end{gather}
and it is a straightforward exercise to verify that 
\eqref{eq_temp2} and \eqref{eq_temp3} do indeed satisfy
the rather strange-looking conservation equation \eqref{eq_temp4}.

Within the Brans-Dicke frame, we may freely set the \emph{entire superfields}
$G$ and $R$ to zero, and then send $\kappa^2$ to zero. Doing so, we find the
supercurrent
\[
J_{\alpha \dalpha} = - \frac{2}{3} \xi \Delta_{\alpha \dalpha} V + W_\alpha W_\dalpha
\]
obeying the conservation equation
\begin{align}
D^\dalpha J_{\alpha \dalpha} = \frac{1}{6} \xi  D_\alpha \bar D^2 V
\end{align}
This is (up to normalizations) the non-covariant conservation equation found
in \cite{Komargodski:2009pc}. It is clear that these non-covariant supercurrents
will yield non-covariant component currents, so we do not bother calculating
those here. They will invariably correspond to the component currents one would
calculate in the Brans-Dicke frame and have no chance of being gauge invariant.

\subsubsection{The Einstein frame}
The choice corresponding to the Einstein frame is to choose the covariantly chiral
superfield $\Phi_0$ to be set to unity. This gauge choice fixes the dilatation
symmetry, while the $U(1)_R$ symmetry is identified with $U(1)_\xi$.
This is easy to prove by examining the covariant chirality condition:
\begin{align}
0 = \nabla^\dalpha \Phi_0 =  D^\dalpha \Phi_0 - \frac{2i}{3} A^\dalpha \Phi_0
     + \frac{i}{3} \kappa^2\xi A^\dalpha_\xi \Phi_0
\end{align}
In the gauge $\Phi_0=1$, this implies $A^\dalpha = \frac{1}{2} \kappa^2 \xi A_\xi^\dalpha$.
A similar argument for the conjugate superfield then necessarily implies that for the
full superfield connections
\begin{gather}
A_\alpha = \frac{1}{2} \kappa^2 \xi A^\xi_\alpha, \;\;\;
A^\dalpha = \frac{1}{2} \kappa^2 \xi A_\xi^\dalpha \eol
A_{\alpha \dalpha} = -\frac{3}{2} G_{\alpha \dalpha} + \frac{1}{2} \kappa^2 \xi A^\xi_{\alpha \dalpha}
\end{gather}
in this conformal gauge.\footnote{The appearance of $G_{\alpha \dalpha}$ in the
bosonic connection has to do with the convention of defining
$F_{\alpha \dalpha} = -3i G_{\alpha \dalpha}$ for the $U(1)$
curvature. Choosing $F_{\alpha \dalpha} = 0$ removes $G_{\alpha \dalpha}$
from the expression for $A_{\alpha \dalpha}$.}

The superfields $\hat G_{\alpha \dalpha}$ and $\hat R$ become the
superfields with the same names of Poincar\'e supergravity. They
obey the constraint
\begin{align}\label{eq_GRconstraint}
\CD_\alpha R - \CD^\dalpha G_{\alpha \dalpha} = X_\alpha
     = -\xi \kappa^2 W_\alpha
\end{align}
where $\CD$ is the Poincar\'e derivative and $X_\alpha$ is the
gaugino superfield associated with the $U(1)_R$ structure, now
identified with $U(1)_\xi$. The conservation equation becomes
\begin{align}
\CD^\dalpha J_{\alpha \dalpha} = -\frac{1}{3} \CD_\alpha J_0
\end{align}
where
\begin{gather}
J_{\alpha \dalpha} = \frac{2}{\kappa^2} G_{\alpha \dalpha} + W_\alpha W_\dalpha, \;\;\; 
J_0 = -\frac{6}{\kappa^2} R
\end{gather}
and follows trivially from \eqref{eq_GRconstraint} and the $U(1)_\xi$
equation of motion.

Observe that all these superfields are manifestly gauge covariant under
the $U(1)_\xi$, which has been absorbed into the structure group of
superspace via its identification with the $U(1)_R$.
It necessarily follows
that any components of these superfields, such as the component
stress tensor or supersymmetry current, ought to share this manifest covariance.

The curious feature of the Einstein frame is that it necessarily
intertwines the $U(1)_\xi$ with supergravity. The instinct that
global supersymmetry should be restored by sending $G_{\alpha \dalpha}$
and $R$ to zero \emph{as superfields} is incorrect in this frame by virtue of
the constraint \eqref{eq_GRconstraint}. One instead suspects that one
must turn off the supergravity multiplet by setting its \emph{component} fields
to zero and then sending $\kappa^2$ to zero. This rather intricate
structure is, unfortunately, the only way to maintain $U(1)_\xi$ covariance.
Proving that this yields the correct globally supersymmetric currents
is the only task left.

\subsection{The component currents}\label{section_ccurrents}
The previous subsection has relied rather heavily on superfield current
arguments. While elegant, this is a somewhat unsatisfying line of attack
since superspace should tell us nothing that we couldn't have already deduced
by more difficult means in component language. We therefore turn to a calculation
of the component currents associated with the superconformal first order
action \eqref{eq_super1var}. In analogy to the gauge supercurrent calculation in
section \ref{review_gauge}, we anticipate that the component first order action
should have the following form:
\begin{align}
e^{-1} \Lag^{(1)} =\, 
     & \delta \phi_0 J_{0}^{(\phi)}
     + \delta \chi^{\alpha}_0 J_{0 \alpha}
     + \delta F_0 J_0 + \hc \eol
     & + \delta A_m{}^\xi J_\xi^m
     + \delta \lambda^{\alpha \xi} J_{\xi \alpha}
     + \delta \bar\lambda_\dalpha^{\xi} J^\dalpha_{\xi}
     + \delta D^\xi J_\xi^{(D)} \eol
     & + \delta e_m{}^a J_a{}^m
     + \delta \psi_m{}^\alpha J_\alpha{}^m
     + \delta \bpsi_m{}_\dalpha J^\dalpha{}^m
     + \delta A_m J^m{}^{(5)}
\end{align}
The components of the chiral compensator multiplet are $(\phi_0, \chi_{\alpha 0}, F_0)$;
those of the $U(1)_\xi$ gauge multiplet are $(A_m^\xi, \lambda_\alpha^\xi, \bar\lambda_\dalpha^\xi, D^\xi)$;
those of the conformal supergravity multiplet
are $(e_m{}^a, \psi_m{}^\alpha, \bpsi_{m \dalpha}, A_m)$. Recall that the last of
these, $A_m$, is the gauge field associated with the $U(1)_R$ of the superconformal
algebra.\footnote{There also exists a gauge field associated with dilatations; however,
its coefficient in the above expansion must vanish (provided the original
action was conformally invariant) since it is the only field which transforms
under special conformal transformations.\cite{butter1}}

Clearly the various component $J$'s described above must be defined in terms
of the superfield expressions considered in the previous section. We have worked
out these relations \cite{butter?} and some are rather complicated. For our purposes,
we will simplify them by ignoring all terms involving the background gravitino.
For the chiral and gauge sectors, the results are exactly those of
global supersymmetry:
\begin{gather}
J_i^{(\phi)} \sim - \frac{1}{4} \nabla^2 J_i, \;\;\;
J_{i \alpha} \sim -\frac{1}{\sqrt 2} \nabla_\alpha J_i
\end{gather}
and for the gauge sector,
\begin{gather}
J_{\xi\alpha} \sim \frac{1}{2} \nabla_\alpha J_\xi, \;\;\;
J_\xi^\dalpha \sim \frac{1}{2} \nabla^\dalpha J_\xi \\
J_{\xi}{}^a \sim \frac{1}{2} \Delta^a J_\xi \\
J_\xi^{(D)} \sim -\frac{1}{2} J_\xi
\end{gather}

For the $U(1)_R$ and supersymmetry currents, we have
\begin{gather}
J^{a (5)} \sim -2 J^a \\
J_{\alpha}{}^b \sim -\frac{i}{2} \bsigma^{b\, \dbeta \beta} J_{\ul{\alpha \beta} \dbeta}
     + \frac{i}{4} \bphi_0 \sigma^b_{\alpha \dalpha} \nabla^\dalpha \bar J_0
\end{gather}
where we have defined
\begin{align}
J_{\ul {\alpha \beta} \dbeta} \equiv \frac{1}{2} \nabla_\alpha J_{\beta \dbeta} + \frac{1}{2} \nabla_\beta J_{\alpha \dbeta}
\end{align}
For the stress tensor,
\begin{align}
J_{ab} \sim & -\frac{1}{2} \Delta_a J_b - \frac{1}{2} \Delta_b J_a + \frac{1}{4} \eta_{ab} \Delta_c J^c \eol
     & - \frac{1}{2} (\chi_0 \sigma_{ab})^\alpha J_{\alpha 0}
     - \frac{1}{2} (\lambda^\xi \sigma_{ab})^\alpha J_{\alpha \xi} + \hc \eol
     & + \frac{1}{4} \eta_{ab}\biggl(
          2 D^\xi J_\xi
          + \frac{3}{2} \lambda^{\alpha \xi} J_{\alpha \xi}
          + \frac{3}{2} \bar\lambda^{\xi}_\dalpha \bar J_\xi^\dalpha
          \biggr) \eol
     & + \frac{1}{4} \eta_{ab}\biggl(
          2 F_0 J_0
          + \frac{3}{2} \chi^\alpha_0 J_{\alpha 0}
          + \phi_0 J_0^{(\phi)}
          + \hc
          \biggr)
\end{align}

These formulae are rather complicated (even after neglecting the gravitino!)
and so it is useful to have a number of independent checks.
The third and fourth lines of $J_{ab}$ correspond to the trace, which is
dictated by scale invariance of the component action. Similarly,
the second line of $J_{ab}$ is antisymmetric and dictated by Lorentz
invariance. The first line of $J_{ab}$ and the entirely of $J_\alpha{}^b$
can be checked using the fermionic special superconformal symmetry.

As these superconformal currents are gauge invariant, fixing the conformal gauge
should yield gauge invariant Einstein frame Poincar\'e currents.\footnote{Note
that Dienes and Thomas have also explored the components of the
superstress tensor in \cite{Dienes:2009td}.}

\subsubsection{The Einstein frame currents}
We begin by putting the entire gauge sector on shell, eliminating the $J_\xi$
multiplet. Then going to the conformal gauge $\Phi_0 = 1$, we find that
the fermion $\chi^\alpha_0$ vanishes and $F_0 = 2 \bar R$
becomes the scalar auxiliary field of Poincar\'e supergravity.
Our goal is to calculate the $U(1)_R$, supersymmetry, and energy-momentum currents
associated with the super-Maxwell system with an FI term. A Noether
current calculation on the globally supersymmetric system (see Appendix \ref{noether} for a brief
discussion) gives
\begin{gather}
J_m^{(5)} = \lambda \sigma_m \bar\lambda \\
J_\alpha{}^m =
     \frac{1}{2} F^{mn} (\sigma_n \lambda)_\alpha
     - \frac{i}{4} \eps^{mdcb}(\sigma_b \bar\lambda)_\alpha F_{cd}
    - \frac{i}{2} \xi (\sigma^m \bar\lambda)_\alpha \\
J_{nm} = F_n{}^p F_{m p} - \frac{1}{4} \eta_{nm} F^{a b} F_{a b}
     + \frac{i}{4} (\lambda \sigma_{\{m} \CD_{n\}} \bar\lambda)
     + \frac{i}{4} (\bar\lambda \bsigma_{\{m} \CD_{n\}} \lambda)
     - \frac{1}{2} \eta_{nm} \xi^2
\end{gather}

We begin with the $U(1)_R$ current:
\begin{align}
J^{a (5)} \sim -2 J^a \sim -\frac{4}{\kappa^2} G^a + \lambda \sigma^a \bar\lambda
\end{align}
The lowest component of $G^a$ belongs to the supergravity multiplet
and so we will set it to zero when decoupling supergravity, giving
identically what the Noether procedure dictates.

For the supersymmetry current, we have
\[
J_{\alpha}{}^b \sim -\frac{i}{2} \bsigma^{b\, \dbeta \beta} J_{\ul{\alpha \beta} \dbeta}
     + \frac{i}{4} \bphi_0 \sigma^b_{\alpha \dalpha} \nabla^\dalpha \bar J_0
\]
and we first need to calculate
\[
J_{\ul {\alpha \beta} \dbeta}
     \equiv \frac{1}{2} \CD_{\{\alpha} J_{\beta\} \dbeta}
     = \frac{1}{\kappa^2} \CD_{\{\alpha} G_{\beta\} \dbeta}
     + \frac{1}{2} \CD_{\{\alpha} W_{\beta\}} W_\dbeta
\]
To evaluate the spinor derivative of $G$ requires certain supergravity relations,
given for example in \cite{bgg}, but all of the terms contained within
involve the gravitino and so vanish when the supergravity sector is turned off.
The second term involves the product of a gaugino and a field strength;
when contracted with a sigma matrix it gives
\[
-\frac{i}{2} \bsigma^{b\, \dbeta \beta} J_{\ul{\alpha \beta} \dbeta}
     = -\frac{1}{2} (\sigma_c \bar\lambda)_\alpha F^{cb}
     - \frac{i}{4} \eps^{dcba} (\sigma_a \bar\lambda)_\alpha F_{cd}
\]
Next we need to calculate the additional
spin-1/2 term involving $\BCD_\dalpha \bar J_0$. Recall that
$\bar J_0 = -6 \bar R / \kappa^2$ and so,
consulting the structure of Poincar\'e superspace \cite{bgg}, one finds that
\[
\BCD_\dalpha \bar J_0 = -\frac{6}{\kappa^2} \BCD_\dalpha \bar R
     \sim \frac{2}{\kappa^2} X_\dalpha
     = -2 \xi W_\dalpha
\]
Thus we find that the supersymmetry current is
\begin{align}
J_\alpha{}^b = -\frac{1}{2} (\sigma_c \bar\lambda)_\alpha F^{cb}
     - \frac{i}{4} \eps^{dcba} (\sigma_a \bar\lambda)_\alpha F_{cd}
     -\frac{i}{2} \xi (\sigma^b \bar \lambda)_\alpha
\end{align}
which is indeed the same (up to index reshufflings) as that produced
by the Noether procedure.

The stress tensor is a significantly more complicated beast. Assuming
the gauge sector is on shell and that the chiral compensator is in
the appropriate gauge, we find the terms
\begin{align}
J_{ab} \sim & -\frac{1}{2} \Delta_a J_b - \frac{1}{2} \Delta_b J_a + \frac{1}{4} \eta_{ab} \Delta_c J^c \eol
     & + \frac{1}{4} \eta_{ab}\biggl(
          2 F_0 J_0
          + J_0^{(\phi)}
          + \hc
          \biggr)
\end{align}
We begin with the calculation of the symmetric traceless part of
\[
\Delta_a J_b = \frac{2}{\kappa^2} \Delta_a G_b - \frac{1}{2} \Delta_a (W\sigma_b \bar W)
\]
The first term yields the symmetric traceless part of the Ricci tensor
and so we can discard it in the limit where we turn off supergravity.
The second term is more complicated and yields
\[
-\frac{i}{4} (\lambda \sigma_{\{b} \nabla_{a\}} \bar\lambda)
               -\frac{i}{4} (\bar \lambda \bsigma_{\{b} \nabla_{a\}} \lambda)
               - F_a{}^m F_{b m}
               + \ldots
\]
where $\ldots$ denotes terms which are either antisymmetric or proportional
to $\eta_{ab}$. Next we calculate the trace of the current. Dropping all
terms which vanish when supergravity is turned off gives
\begin{align}
\frac{1}{4} J_0^{(\phi)} + \hc
     \ni \frac{3}{8 \kappa^2} (\CD^2 R + \BCD^2 \bar R)
     \ni -\frac{1}{4 \kappa^2} \CD^\alpha X_\alpha
     = + \frac{\xi}{4} \CD^\alpha W_\alpha
     = -\frac{1}{2} \xi^2
\end{align}
Putting everything together, we find the stress energy tensor
\begin{align}
J_{ab} = F_a{}^m F_{b m} - \frac{1}{4} \eta_{ab} F^{mn} F_{mn}
          +\frac{i}{4} (\lambda \sigma_{\{b} \partial_{a\}} \bar\lambda)
          +\frac{i}{4} (\bar \lambda \bsigma_{\{b} \partial_{a\}} \lambda)
          -\frac{1}{2} \xi^2 \eta_{ab}
\end{align}
This, too, is as expected from the Noether procedure.

\section{Conclusion}
We have attempted to demonstrate that the curious features of conventional (old minimal)
supergravity make defining the supercurrent superfield somewhat subtle.
The most straightforward way of going about it involves
a Brans-Dicke frame which for the case of an FI term yields an inconsistent
energy momentum tensor, while the current analogous to the Einstein frame
involves a subtle intertwining of gravity and gauge fields prior to
the decoupling of supergravity. In particular, the supersymmetry
generators $Q$ now carry $U(1)_\xi$ charge, and thus so does the
gravitino and any superpotential we turn on. The difficulties
with the latter have been discussed particularly in
\cite{Komargodski:2009pc, Dienes:2009td}
and references therein and remains a major objection to any FI term
in conventional supergravity. Moreover, that the gravitino should
carry some $U(1)_\xi$ charge implies that maintaining gauge invariance
at the \emph{quantum} level might necessitate some nontrivial
anomaly cancellation.

Other flavors of supergravity can accommodate FI terms more easily.
New minimal supergravity, for example, involves a different supercurrent conservation equation
as it involves a linear superfield $L$ as a conformal compensator \cite{Buchbinder:1998qv}.
In the superconformal framework, FI terms in such a theory take the
simple form
\[
2\xi \int d^8z\, E\, L V
\]
where $L$ is conformal dimension two. Gauge invariance holds since
the integral of $L \Lambda$ for chiral $\Lambda$ vanishes.
The Poincar\'e theory is recovered by taking $L$ to unity, and the
FI term has exactly the same form in the supergravity theory as it
does in the globally supersymmetric one. The difficulty with
new minimal supergravity is that it does not fix the $U(1)_R$
invariance, which remains a symmetry of the Poincar\'e theory. This
severely constrains the type of matter which can be coupled
even at the classical level, while at the quantum level
it is generically anomalous \cite{superspace, Gates:1981yc}.

It is a possibility that one could combine features of both theories by
using both chiral and linear compensators, which would yield a
non-minimal supergravity with sixteen off-shell bosonic and sixteen
off-shell fermionic degrees of freedom, but could more straightforwardly
accommodate FI terms (and K\"ahler potentials). Indeed, this seems like it could
lie at the root of the $S$-multiplet recently constructed by Komargodski
and Seiberg \cite{Komargodski:2010rb} and warrants further investigation.

\newpage
\appendix
\section{Noether currents for super-Maxwell theory}\label{noether}
The component super-Maxwell theory (with FI term) has the Lagrangian
\begin{align}
\Lag = - \frac{1}{4} F^{mn} F_{mn}
     - \frac{i}{2} \bar \lambda \bsigma^m \CD_m \lambda
     - \frac{i}{2} \lambda \sigma^m \CD_m \lambda
     + \frac{1}{2} D^2 - \xi D
\end{align}
and is invariant (up to a total derivative) under the global supersymmetry
transformations
\begin{gather}
\delta_\xi D = i (\xi \sigma^m \CD_m \bar\lambda) + \hc \\
\delta_\xi \lambda^\alpha = -i (\xi \sigma^{ba})^\alpha F_{ab} + \xi^\alpha D \\
\delta_\xi A_{m} = \xi \sigma_m \bar\lambda + \hc
\end{gather}
where the spinor parameter $\xi^\alpha$ is not to be confused with the
FI parameter $\xi$.

A comment is necessary about this definition for the supersymmetry
transformations. While perfectly sensible in the component formulation, these
transformations do not correspond to the most straightforward definition of
``supersymmetry'' as supertranslation in superspace. Rather, if
one works with conventionally chiral superfields and Wess-Zumino gauge
for the $U(1)$ prepotential, one must augment the supertranslation by 
a further gauge transformation to restore Wess-Zumino gauge, which
yields these rules. If instead
one works with covariantly chiral superfields, the supersymmetry
transformation described above is exactly that of a covariant super Lie
derivative in superspace.\footnote{See the discussion in \cite{bgg}.}
This latter interpretation is clearly more elegant.

We define the supersymmetry current $J_{\alpha}{}^m$ by calculating
the shift in the action for local $\xi^\alpha$ when the gauge sector
is on shell. One finds
\begin{align}
\delta_\xi S = -2 \int d^4x\, \partial_m \xi^\alpha J_\alpha{}^m
\end{align}
where
\begin{align}
J_\alpha{}^m =
     \frac{1}{2} F^{mn} (\sigma_n \lambda)_\alpha
     - \frac{i}{4} \eps^{mdcb}(\sigma_b \bar\lambda)_\alpha F_{cd}
    - \frac{i}{2} \xi (\sigma^m \bar\lambda)_\alpha
\end{align}
which is manifestly gauge invariant. The factor of two in the
definition is conventional since under the normalization of
supersymmetry used here, the gravitino transforms as
$2 \partial_m \xi^\alpha$.

A caveat is also necessary regarding the construction of the
current associated with translations. As is well known, the
energy-momentum tensor for physical reasons ought to be both symmetric and
gauge invariant, but the straightforward Noether current constructed from
translations is neither, and must be augmented by a certain ``improvement term,''
whose addition changes neither the conservation of the current nor the
definition of the conserved charge. The result, known as the
Belinfante tensor, is conserved, symmetric, and gauge invariant
(see for example the discussion in \cite{Weinberg:1995mt}).
For the super-Maxwell Lagrangian here, the Belinfante tensor is
\begin{align}
J_{nm} = F_n{}^p F_{m p} - \frac{1}{4} \eta_{nm} F^{a b} F_{a b}
     + \frac{i}{4} (\lambda \sigma_{\{m} \CD_{n\}} \bar\lambda)
     + \frac{i}{4} (\bar\lambda \bsigma_{\{m} \CD_{n\}} \lambda)
     - \frac{1}{2} \eta_{nm} \xi^2
\end{align}
Note that its trace is proportional to
$\xi^2$ (since the gauginos obey their equation of motion
$\bsigma^m \partial_m \lambda$), and so it is only the
FI term which spoils the classical scale invariance of the theory.

In addition to global supersymmetry and translations, the theory described
above admits a global $U(1)_R$ rotation under which the supersymmetry
generator $Q_\alpha$ has charge -1 and $Q^\dalpha$ has charge +1.
This leads to $\lambda_\alpha$ carrying charge $+1$, with the other fields
neutral, and so the $U(1)_R$ current is simply the gaugino vector bilinear
\begin{align}
J_m^{(5)} = \lambda \sigma_m \bar\lambda
\end{align}
where we have defined
\begin{align}
\delta S = -\int d^4x\, \partial_m \alpha_{(5)} J^{m (5)}
\end{align}
for local $U(1)_R$ rotation parameter $\alpha_{(5)}$.

It is a straightforward exercise to check that each of these Noether
currents is conserved using their equations of motion, and they
agree with the currents constructed in \cite{Dienes:2009td}
(up to improvement terms).


\section*{Acknowledgments}
I am grateful to Mary K. Gaillard for useful discussions as well as
Zohar Komargodski, Keith Dienes, and Brooks Thomas for clarifying remarks on
earlier drafts.
This work was supported in part by the Director, Office of Science, Office of High Energy and Nuclear Physics, Division of High Energy Physics of the U.S. Department of Energy under Contract DE-AC02-05CH11231, in part by the National Science Foundation under grant PHY-0457315.


\end{document}

%% file: sugracom.tex
\newcommand{\beq}{\begin{equation}}
\newcommand{\eeq}{\end{equation}}
\newcommand{\ul}{\underline}

\newcommand{\eps}{{\epsilon}}
\newcommand{\abs}[1]{\left| #1 \right|}

\newcommand{\hc}{\mathrm{h.c.}}
\newcommand{\Lag}{\mathcal L}

\newcommand{\CD}{\mathcal{D}}
\newcommand{\BCD}{\mathcal{\bar D}}

\newcommand{\CP}{{\mathcal P}}
\newcommand{\ACP}{{\bar {\mathcal P}}}

\newcommand{\bsigma}{\bar{\sigma}}
\newcommand{\dalpha}{{\dot{\alpha}}}
\newcommand{\dbeta}{{\dot{\beta}}}

\newcommand{\btheta}{{\bar\theta}}

\newcommand{\chE}{\mathcal E}